\documentclass{aip-cp}
\usepackage{amsfonts}
\usepackage[numbers]{natbib}
\usepackage{rotating}
\usepackage{graphicx}
\newcommand\rf[1]{(\ref{eq:#1})}
\newcommand\lab[1]{\label{eq:#1}}
\newcommand\nonu{\nonumber}
\newcommand\br{\begin{eqnarray}}
\newcommand\er{\end{eqnarray}}
\newcommand\be{\begin{equation}}
\newcommand\ee{\end{equation}}

\newcommand\lb{\lbrack}
\newcommand\rb{\rbrack}

\newcommand\llb{\left\lbrack}
\newcommand\rrb{\right\rbrack}

\renewcommand\({\left(}
\renewcommand\){\right)}
\newcommand\bgv{\bigg\vert}              

\newcommand\bc{\begin{center}}
\newcommand\ec{\end{center}}

\renewcommand\a{\alpha}

\renewcommand\d{\delta}
\newcommand\eps{\epsilon}
\newcommand\vareps{\varepsilon}

\newcommand\G{\Gamma}

\newcommand\h{\frac{1}{2}}
\renewcommand\k{\kappa}
\renewcommand\l{\lambda}
\renewcommand\L{\Lambda}
\newcommand\m{\mu}
\newcommand\n{\nu}
\newcommand\om{\omega}

\newcommand\vp{\varphi}
\renewcommand\P{\Phi}
\newcommand\pa{\partial}

\newcommand\pr{\prime}

\newcommand\s{\sigma}

\renewcommand\t{\tau}

\newcommand\wti{\widetilde}

\newcommand\cA{{\mathcal A}}
\newcommand\cB{{\mathcal B}}

\newcommand\cF{{\mathcal F}}

\newcommand\cH{{\mathcal H}}

\newcommand\cL{{\mathcal L}}
\newcommand\cM{{\mathcal M}}

\newcommand\cU{{\mathcal U}}

\newcommand{\ct}[1]{\cite{#1}}
\newcommand\PRL[3]{{\em Phys. Rev. Lett.} \textbf{#1}, #3 (#2)}

\newcommand\PRD[3]{{\em Phys. Rev.} \textbf{D#1}, #3 (#2)}

\newcommand\PLB[3]{{\em Phys. Lett.} \textbf{#1B}, #3 (#2)}

\newcommand\AoP[3]{{\em Ann. of Phys.} \textbf{#1}, #3 (#2)}

\newcommand\IJMPA[3]{{\em Int. J. Mod. Phys.} \textbf{A#1}, #3 (#2)}
\newcommand\IJMPD[3]{{\em Int. J. Mod. Phys.} \textbf{D#1}, #3 (#2)}

\newcommand\MPLA[3]{{\em Mod. Phys. Lett.} \textbf{A#1}, #3 (#2)}

\newcommand\udot{\stackrel{.}{u}}

\newcommand\Adot{\stackrel{.}{A}}
\newcommand\Bdot{\stackrel{.}{B}}
\newcommand\Hdot{\stackrel{.}{H}}

\begin{document}

\title{Modified Gravity and Inflaton Assisted Dynamical Generation 
of Charge Confinement and Electroweak Symmetry Breaking in 
Cosmology}

\author[aff1,aff2,aff3]{Eduardo Guendelman 
}
\eaddress{guendel@bgu.ac.il}
\author[aff4]{Emil Nissimov\corref{cor1} 
}
\author[aff4]{Svetlana Pacheva 
}
\eaddress{svetlana@inrne.bas.bg}

\affil[aff1]{Physics Department, Ben Gurion University of the Negev, Beer Sheva, 
Israel}
\affil[aff2]{Bahamas Advanced Study Institute and Conferences, 4A Ocean Heights, 
Hill View Circle, Stella Maris, Long Island, The Bahamas}
\affil[aff3]{Frankfurt Institute for Advanced Studies, Giersch Science Center, 
Campus Riedberg, Frankfurt am Main, Germany}
\affil[aff4]{Institute for Nuclear Research and Nuclear Energy, Bulgarian Academy 
of Sciences, Sofia, Bulgaria}

\corresp[cor1]{Corresponding author: nissimov@inrne.bas.bg}

\maketitle

\begin{abstract}
We describe a new type of gravity-matter models where modified $f(R)=R+R^2$
gravity couples non-canonically to a scalar ``inflaton'', to the bosonic sector 
of the electroweak particle model and to a special nonlinear gauge field with 
a square-root of the standard Maxwell/Yang-Mills kinetic term simulating 
QCD confining dynamics. Our construction is based  on the powerful formalism 
of non-Riemannian space-time volume-forms -- alternative metric-independent
volume elements defined in terms of auxiliary antisymmetric tensor gauge 
fields. 
Our model provides a unified Lagrangian action principle description of: 
(i) the evolution of both ``early'' and ``late'' universe by the ``inflaton''
scalar field; 
(ii) gravity-inflaton-assisted dynamical generation of Higgs spontaneous 
breakdown of electroweak gauge symmetry in the ``late'' universe, as well as 
dynamical suppression of electroweak breakdown in the ``early'' universe;
(iii) gravity-inflaton-assisted dynamical generation of QCD-like 
confinement in the ``late'' universe and suppression of confinement in 
the ``early'' universe due to the special interplay with the dynamics of 
the QCD-simulating nonlinear gauge field.
\end{abstract}

\section{1. Introduction}
\label{intro}

One of the principal tasks in cosmology is the establishment from first
principles, \textsl{i.e.}, from Lagrangian action principle, of consistent
mechanisms driving the appearance, respectively the suppression, of confinement 
and electroweak spontaneous symmetry breaking during the various stages in
the evolution of the universe \ct{general-cit-1}-\ct{general-cit-7}.

In the present note we will discuss in some detail the main interesting
properties of a new type of non-canonical extended gravity-matter model, in
particular, its implications for cosmology.
Namely we will consider modified  $f(R)=R+R^2$ gravity coupled in a
non-standard way to a scalar ``inflaton'' field, to the bosonic fields of
the standard electroweak particle model, as well as to a special kind of a
nonlinear (Abelian or non-Abelian) gauge field with a square-root of the 
standard Maxwell/Yang-Mills kinetic term which simulates QCD confining dynamics.
In this way our model will represent qualitatively extended gravity coupled
to the whole (bosonic part of the) standard model of elementary particle
physics.

The first essential non-standard feature of the model under consideration is
its construction in terms of non-Riemannian spacetime volume-forms
(alternative metric-independent generally covariant volume elements) defined in
terms of auxiliary antisymmetric tensor gauge fields of maximal rank
(see Refs.\ct{susyssb-1,grav-bags} for a consistent geometrical formulation,
which is an extension of the originally proposed method
\ct{TMT-orig-1,TMT-orig-2}).
The latter volume-form gauge fields were shown to be almost pure-gauge -- 
apart from few arbitrary 
integration constants they do not produce additional  propagating field-theoretic 
degrees of freedom (see Appendices A of Refs.\ct{grav-bags,grf-essay} and 
Section 2 below).
Yet the non-Riemannian spacetime volume-forms trigger a series of important 
physical features unavailable in ordinary gravity-matter models with the 
standard Riemannian volume element (given by the square-root of the determinant 
of the Riemannian metric):

(i) The ``inflaton'' $\vp$ develops a remarkable effective scalar potential in
the Einstein frame possessing an infinitely large flat region for large
negative $\vp$ describing the ``early'' universe evolution; 

(ii) In the absence of the $SU(2)\times U(1)$ iso-doublet scalar field, 
the ``inflaton'' effective potential has another infinitely large flat region 
for large positive $\vp$ at much lower energy scale describing the ``late'' 
post-inflationary (dark energy dominated) universe;

(iii) Inclusion of the $SU(2)\times U(1)$ iso-doublet scalar field $\s$ --
{\em without the usual tachyonic mass and quartic self-interaction term}
-- introduces a drastic change in the total effective scalar potential in the
post-inflationary universe: the effective potential as a function of $\s$ 
dynamically acquires exactly the electroweak Higgs-type spontaneous symmetry 
breaking form. The latter is an explicit realization of Bekenstein's idea
\ct{gravity-assist-86} for a gravity-assisted dynamical electroweak spontaneous 
symmetry breaking.

(iv) Further important features arise upon introducing a coupling to an additional
strongly nonlinear gauge field whose Lagrangian contains a square-root of
the standard Maxwell/Yang-Mills kinetic term. The latter is known to
describe charge confinement in flat spacetime \ct{GG-2} as well as 
in curved spacetime for static spherically symmetric field configurations 
(Appendix B in Ref.\ct{grav-bags}; see also Eq.\rf{cornell-type} below). 
This is a simple implementation of `t Hooft's idea \ct{thooft} about confinement
being produced due to the presence in the energy density of electrostatic 
field  configurations of a term {\em linear} w.r.t. electric displacement field 
in the infrared region (arising presumably as an appropriate infrared counterterm).
Therefore, the addition of the ``square-root'' nonlinear gauge field will simulate 
the strong interactions QCD-like dynamics. 

Let us particularly emphasize that the specific form of the action (Eq.\rf{TMMT-1}
below) with the several non-Riemannian volume elements describing our model is 
uniquely determined by the requirement of global Weyl-scale symmetry
(Eq.\rf{scale-transf} below) which becomes spontaneously broken upon 
transferring to the physical Einstein frame.

As a result, in the Einstein frame we achieve:

(a) Bekenstein-inspired gravity-inflaton-assisted dynamical generation of Higgs-type 
electroweak spontaneous symmetry breaking in the ``late'' universe, while there 
is no electroweak breaking in the ``early'' universe;

(b) Simultaneously we obtain gravity-inflaton-assisted dynamical generation of charge 
confinement in the ``late'' universe as well as gravity-suppression of confinement,
\textsl{i.e.}, deconfinement in the ``early'' universe.

In Section 2 we briefly review the main properties of the non-Riemannian 
volume-forms on spacetime manifolds, including elucidating the (almost) pure
gauge nature of the associated antisymmetric tensor gauge fields of maximal
rank.
In Section 3 we first provide the formulation of our non-canonical $f(R)$-gravity
model coupled to the (bosonic part of the) standard model of elementary particles
in terms of several  different non-Riemannian spacetime volume-forms. Next we
describe the construction of the corresponding physical Einstein-frame action.
Section 4 discusses the main interesting implications for cosmology of the
present model. Section 5 contains some conclusions and outlook.


\section{2. Non-Riemannian Volume-Forms in Gravitational Theories}
\label{nonriemannian}
\subsubsection{2.1 Non-Riemannian Volume-Forms - General Properties}
\label{nonrriemannian-gen}
Volume-forms (generally-covariant integration measures) in integrals over
manifolds are given by nonsingular maximal rank differential forms $\om$:
\br
\int_{\cM} \om \bigl(\ldots\bigr) = \int_{\cM} dx^D\, \Omega \bigl(\ldots\bigr)
\;\; ,\;\; 
\om = \frac{1}{D!}\om_{\m_1 \ldots \m_D} dx^{\m_1}\wedge \ldots \wedge dx^{\m_D}\; ,
\lab{omega-1} \\
\om_{\m_1 \ldots \m_D} = - \vareps_{\m_1 \ldots \m_D} \Omega \;\; ,\;\;
dx^{\m_1}\wedge \ldots \wedge dx^{\m_D} = \vareps^{\m_1 \ldots \m_D}\,  dx^D \; ,
\lab{omega-3}
\er
(our conventions for the alternating symbols $\vareps^{\m_1,\ldots,\m_D}$ and
$\vareps_{\m_1,\ldots,\m_D}$ are: $\vareps^{01\ldots D-1}=1$ and
$\vareps_{01\ldots D-1}=-1$).
The volume element (integration measure density) $\Omega$ transforms as scalar
density under general coordinate reparametrizations.

In standard generally-covariant theories (with action $S=\int d^D\! x \sqrt{-g} \cL$)
the Riemannian spacetime volume-form is defined through the ``D-bein''
(frame-bundle) canonical one-forms $e^A = e^A_\m dx^\m$ ($A=0,\ldots ,D-1$):
\be
\om = e^0 \wedge \ldots \wedge e^{D-1} = \det\Vert e^A_\m \Vert\,
dx^{\m_1}\wedge \ldots \wedge dx^{\m_D} \quad  \longrightarrow \quad
\Omega = \det\Vert e^A_\m \Vert\, d^D x = \sqrt{-\det\Vert g_{\m\n}\Vert}\, d^D x \; .
\lab{omega-riemannian}
\ee

There is {\em no a priori} any obstacle to employ instead of $\sqrt{-g}$ another
alternative {\em non-Riemannian} volume element as in \rf{omega-1}-\rf{omega-3}
given by a non-singular {\em exact} $D$-form $\om = d B$ where:
\be
B = \frac{1}{(D-1)!} B_{\m_1\ldots\m_{D-1}} 
dx^{\m_1}\wedge\ldots\wedge dx^{\m_{-1}} \; ,
\lab{B-form}
\ee
so that the {\em non-Riemannian} volume element reads:
\be
\Omega \equiv \Phi(B) = 
\frac{1}{(D-1)!}\vareps^{\m_1\ldots\m_D}\, \pa_{\m_1} B_{\m_2\ldots\m_D} \; .
\lab{Phi-D}
\ee

Here $B_{\m_1\ldots\m_{D-1}}$ is an auxiliary rank $(D-1)$ antisymmetric tensor gauge
field. $\Phi(B)$, which is in fact the density of the dual of the rank $D$ field strength
$F_{\m_1 \ldots \m_D} = \frac{1}{(D-1)!} \pa_{\lb\m_1} B_{\m_2\ldots\m_D\rb}
= - \vareps_{\m_1 \ldots \m_D} \P (B)$, 
similarly transforms as scalar density under general coordinate reparametrizations.

The presence of non-Riemannian volume element $\Phi(B)$ in a gravity-matter action
$S=\int d^D\! x \Phi(B) \cL + \ldots$ {\em does not} change the number of
field-theoretic degrees of freedom -- the latter remains the same as with the standard
Riemannian measure $\sqrt{-g}$. 

In fact, as we will demonstrate in the next Subsection 2.2, the canonical Hamiltonian 
analysis reveals that the auxiliary gauge field  $B_{\m_1\ldots\m_{D-1}}$ is
(almost) pure-gauge! This is because the total Lagrangian is only linear
w.r.t. $B$-velocities, so it leads to Hamiltonian constraints a'la Dirac.
The only remnant of $B_{\m_1\ldots\m_{D-1}}$ is a {\em discrete degree of freedom} 
which appears as integration constant $M$ in the equations of motion w.r.t. 
$B_{\m_1\ldots\m_{D-1}}$ (see subsect. 3.2 below). 
$M$ is in fact a conserved Dirac constrained canonical momentum
conjugated to the ``magnetic'' $B$-component 
$\frac{1}{(D-1)!}\vareps^{i_1\ldots i_{D-1}} B_{i_1\ldots i_{D-1}}$.

\subsubsection{2.2 Canonical Hamiltonian Treatment of Gravity-Matter Theories with
Non-Riemannian Volume-Forms}
\label{nonrriemannian-hamil}
Here we provide a brief discussion of the application of the canonical Hamiltonian 
formalism to a general gravity-matter model involving several non-Riemannian spacetime
volume elements of the type \rf{TMMT-1} discussed below (see also Appendices A in 
Refs.\ct{grav-bags,grf-essay}):
\br
S = \int d^4 x \P_1 (A) {\wti L}^{(1)} (u,\udot) + 
\int \P_2 (B) \llb {\wti L}^{(2)} (u,\udot) + \frac{\P_4 (H)}{\sqrt{-g}}\rrb \; ,
\lab{TMMT+GG-0} \\
\P_1 (A) = \frac{1}{3!}\vareps^{\m\n\k\l} \pa_\m A_{\n\k\l}  \quad ,\quad
\P_2 (B) = \frac{1}{3!}\vareps^{\m\n\k\l} \pa_\m B_{\n\k\l} \quad ,\quad
\P_4 (H) = \frac{1}{3!}\vareps^{\m\n\k\l} \pa_\m H_{\n\k\l} \; ,
\lab{Phi-1-2-3}
\er
where  the Lagrangians ${\wti L}^{(1,2)}(u,\udot)$ include both matter and scalar curvature terms, and
where $(u,\udot)$ collectively denote the set of the basic gravity-matter canonical variables 
$(u)=\bigl(g_{\m\n},\mathrm{matter}\bigr)$ and their respective velocities.
In \rf{TMMT+GG-0} $\P_4 (H)$ is the density dual of the gauge-field strength of an additional auxiliary gauge field $H_{\m\n\l}$ necessary for the consistency of the model.

For the present purpose it is sufficient to concentrate only on the canonical 
Hamiltonian structure related to the auxiliary maximal rank antisymmetric tensor gauge fields 
$A_{\m\n\l}, B_{\m\n\l}, H_{\m\n\l}$ and their respective conjugate momenta.

For convenience we introduce the following short-hand notations for the dual
field-strengths \rf{Phi-1-2-3} of the auxiliary 3-index antisymmetric gauge 
fields (the dot indicating time-derivative): 
\br
\P_1 (A) = \Adot + \pa_i A^i \quad, \quad 
A = \frac{1}{3!} \vareps^{ijk} A_{ijk} \;\; ,\;\;
A^i = - \h \vareps^{ijk} A_{0jk} \; ,
\lab{A-can} \\
\P_2 (B) = \Bdot + \pa_i B^i \quad, \quad 
B = \frac{1}{3!} \vareps^{ijk} B_{ijk} \;\; ,\;\;
B^i = - \h \vareps^{ijk} B_{0jk} \; ,
\lab{B-can} \\
\P_4 (H) = \Hdot + \pa_i H^i \quad, \quad 
H = \frac{1}{3!} \vareps^{ijk} H_{ijk} \;\; ,\;\;
H^i = - \h \vareps^{ijk} H_{0jk} \; ,
\lab{H-can}
\er

For the pertinent canonical momenta conjugated to \rf{A-can}-\rf{H-can} we have:
\br
\pi_A = {\wti L}^{(1)} (u,\udot) \quad ,\quad
\pi_B = {\wti L}^{(2)} (u,\udot) + \frac{1}{\sqrt{-g}}(\Hdot + \pa_i H^i)
\quad ,\quad  \pi_H = \frac{1}{\sqrt{-g}}(\Bdot + \pa_i B^i) \; ,
\lab{can-momenta-aux}
\er
and:
\be
\pi_{A^i} = 0 \quad,\quad \pi_{B^i} = 0 \quad,\quad \pi_{H^i} = 0 \; .
\lab{can-momenta-zero}
\ee
The latter imply that $A^i, B^i, H^i$ will in fact appear as Lagrange multipliers
for certain first-class Hamiltonian constraints 
(see Eqs.\rf{pi-A-const}-\rf{pi-B-pi-H-const} below). 
For the canonical momenta conjugated to the basic gravity-matter canonical 
variables we have (using last relation \rf{can-momenta-aux}):
\be
p_u = (\Adot + \pa_i A^i) \frac{\pa}{\pa \udot} {\wti L}^{(1)} (u,\udot) + 
\pi_H \sqrt{-g} \frac{\pa}{\pa \udot} {\wti L}^{(2)} (u,\udot) \; .
\lab{can-momenta-u}
\ee

Now, relations \rf{can-momenta-aux} and \rf{can-momenta-u} allow us to
obtain the velocities $\udot,\,\Adot,\,\Bdot,\,\Hdot$ as functions
of the canonically conjugate momenta $\udot = \udot (u,p_u,\pi_A,\pi_B,\pi_H)$
\textsl{etc.} (modulo some Dirac constraints among the basic gravity-matter
variables due to general coordinate and gauge invariances). Taking into account
\rf{can-momenta-aux}-\rf{can-momenta-zero} (and the short-hand notations
\rf{A-can}-\rf{H-can}) the canonical Hamiltonian corresponding to 
\rf{TMMT+GG-0}:
\be
\cH = p_u \udot + \pi_A \Adot + \pi_B \Bdot + \pi_H \Hdot -
(\Adot + \pa_i A^i) {\wti L}_1 (u,\udot) 
- \pi_H \sqrt{-g} \Bigl\lb {\wti L}^{(2)}(u,\udot) + 
\frac{1}{\sqrt{-g}}(\Hdot + \pa_i H^i) \Bigr\rb
\lab{can-hamiltonian}
\ee
acquires the following form as function of the canonically conjugated variables
(here $\udot = \udot (u,p_u,\pi_A,\pi_B,\pi_H)$):
\be
\cH = p_u \udot - \pi_H \sqrt{-g} {\wti L}^{(2)}(u,\udot)
+ \sqrt{-g} \pi_H \pi_B - \pa_i A^i \pi_A - \pa_i B^i \pi_B - \pa_i H^i \pi_H \; .
\lab{can-hamiltonian-final}
\ee
From \rf{can-hamiltonian-final} we deduce that indeed $A^i, B^i, H^i$ are 
Lagrange multipliers for the first-class Hamiltonian constraints:
\be
\pa_i \pi_A = 0 \;\; \to\;\; \pi_A = - M_1 = {\rm const} \; ,
\lab{pi-A-const}
\ee
and similarly:
\be
\pi_B = - M_2 = {\rm const} \quad ,\quad \pi_H = \chi_2 = {\rm const} \; ,
\lab{pi-B-pi-H-const}
\ee
which are the canonical Hamiltonian counterparts of Lagrangian constraint
equations of motion derived in the next Section (see 
\rf{integr-const-1}-\rf{integr-const-3-4} below).

Thus, the canonical Hamiltonian treatment of \rf{TMMT+GG-0} reveals the meaning
of the auxiliary 3-index antisymmetric tensor gauge fields
$A_{\m\n\l},\, B_{\m\n\l},\, H_{\m\n\l}$ -- building blocks of
the non-Riemannian spacetime volume-form formulation of the modified gravity-matter
model \rf{TMMT+GG-0}. Namely, the canonical momenta $\pi_A,\, \pi_B,\, \pi_H$ 
conjugated to the ``magnetic'' parts $A,B,H$ \rf{A-can}-\rf{H-can}
of the auxiliary 3-index antisymmetric tensor gauge fields are constrained
through Dirac first-class constraints \rf{pi-A-const}-\rf{pi-B-pi-H-const}
to be constants identified with the arbitrary 
integration constants $\chi_2,\, M_1,\, M_2$  arising within the 
Lagrangian formulation of the model (see \rf{integr-const-1}-\rf{integr-const-3-4}
below). 
The canonical momenta 
$\pi_A^i,\, \pi_B^i,\, \pi_H^i$ conjugated to the ``electric'' parts $A^i,B^i,H^i$ 
\rf{A-can}-\rf{H-can} of the auxiliary 3-index antisymmetric tensor gauge field
are vanishing \rf{can-momenta-zero} which makes the latter canonical Lagrange 
multipliers for the above Dirac first-class constraints.

\section{3. Non-Canonical $f(R)$-Gravity Model in Terms of Non-Riemannian Spacetime
Volume-Forms}
\label{TMMT}

\subsection{3.1 General Construction}
\label{einstein-frame-gen}

We start with the following non-canonical $f(R)=R+R^2$ gravity-matter action 
constructed in terms of three different non-Riemannian volume-forms (generally
covariant metric-independent volume elements) coupled to an ``inflaton'' and
an additional auxiliary scalar field, as well as to a confining
nonlinear gauge field simulating QCD dynamics and to the bosonic 
sector of the electroweak standard model. The corresponding action, 
generalizing the actions in Refs.\ct{grav-bags,grf-essay,bekenstein} 
reads (for simplicity we use units with the Newton constant $G_N = 1/16\pi$):

\br
S = \int d^4 x\,\P_1(A) \Bigl\lb R + L^{(1)}(\vp,\s,\cA_\m,\psi)\Bigr\rb
+ \int d^4 x\,\P_2(B) \Bigl\lb \eps R^2 + L^{(2)}(\vp,\s,\cA_\m,\psi) +
\frac{\P_4 (H)}{\sqrt{-g}}\Bigr\rb - \int d^4 x\,\P_3(C) \psi^2 \; .
\lab{TMMT-1}
\er
Here the following notations are used:

(i) $\P_1(A)$ and $\P_2(B)$ are the two independent non-Riemannian volume 
elements as in \rf{Phi-1-2-3}, $\P_4(H)$ is the same as in the last 
relation \rf{Phi-1-2-3} and it is needed for consistency of \rf{TMMT-1}. 
Here we introduced also a third independent non-Riemannian volume element 
for the reasons explained after Eq.\rf{} below. 

(ii) We particularly emphasize that we start within the first-order 
{\em Palatini formalism} for the scalar curvature $R$ and the Ricci tensor
$R_{\m\n}$: $R=g^{\m\n} R_{\m\n}(\G)$, where $g_{\m\n}$, $\G^\l_{\m\n}$ --
the metric and affine connection are {\em apriori} independent.

(iii) The first matter field Lagrangian $L^{(1)}(\vp,\s,\cA_\m,\psi)$ in
\rf{TMMT-1} is a sum of ``inflaton''  $L_1 (\vp,X)$  Lagrangian, nonlinear 
$\cA_\m$ gauge field term and the Lagrangian $L_2 (\s,Y)$ of a complex
$SU(2)\times U(1)$ iso-doublet Higgs-like scalar 
$\s \equiv (\s_a)$ coupled to an auxiliary scalar field $\psi$:
\be
L^{(1)}(\vp,\s,\cA_\m,\psi) \equiv L_1 (\vp,X) - \h f_0 \sqrt{-F^2}
+ L_2 (\s,Y;\psi)\; ,
\lab{L-1}
\ee
Here we have explicitly:
\be
L_1 (\vp,X) = X - f_1 e^{-\a\vp} \;\; ,\;\;
X \equiv - \h g^{\m\n} \pa_\m \vp \pa_\n \vp \; ,
\lab{L-1a}
\ee
where $\a, f_1$ are dimensionful positive parameters.
\br
F^2 \equiv F_{\m\n} F_{\k\l} g^{\m\k} g^{\n\l} 
\quad, \quad F_{\m\n} = \pa_\m \cA_\n - \pa_\n \cA_\m \quad
\Bigl( + \lb \cA_\m ,\cA_\n \rb\Bigr)
\nonu
\er
($\cA_\m$ could be either Abelian or non-Abelian, see the discussion below).

$\s \equiv (\s_a)$ is a complex $SU(2)\times U(1)$ iso-doublet Higgs-like 
scalar field with Lagrangian:
\be
L_2 (\s,Y;\psi) = Y - \psi^2 \s^{*}_a \s_a \;\; ,\;\; 
Y \equiv - g^{\m\n}(\nabla_\m \s)^{*}_a \nabla_\n \s_a \; ,
\lab{L-1b}
\ee
where the gauge-covariant derivative acting on $\s$ reads:
\be
\nabla_\m \s = 
\Bigl(\pa_\m - \frac{i}{2} \t_A \cA_\m^A - \frac{i}{2} \cB_\m \Bigr)\s \; ,
\lab{cov-der}
\ee
with $\h \t_A$ ($\t_A$ -- Pauli matrices, $A=1,2,3$) indicating the $SU(2)$ 
generators and $\cA_\m^A \equiv \vec{\cA}$ ($A=1,2,3$)
and $\cB_\m$ denoting the corresponding electroweak $SU(2)$ and $U(1)$ 
gauge fields.

(iv) The second matter field Lagrangian $L^{(2)}(\vp,\s,\cA_\m,\psi)$ in
\rf{TMMT-1} is a sum of the standard Maxwell and Yang-Mills kinetic terms for
$\cA_\m$ and the electroweak gauge fields $(\vec{\cA},\cB)$ and the kinetic 
term for the auxiliary scalar $\psi$:
\be
L^{(2)}(\vp,\s,\cA_\m,\psi) = 
\frac{1}{4e^2} F^2 - \frac{1}{4g^2} \cF^2(\vec{\cA}) - \frac{1}{4g^{\pr\,2}} \cF^2(\cB)
- \h g^{\m\n} \pa_\m \psi \pa_\n \psi \; ,
\lab{L-2}
\ee
where (all $SU(2)$ indices $A,B,C = (1,2,3)$):
\br
\cF^2(\vec{\cA}) \equiv 
\cF^A_{\m\n} (\vec{\cA}) \cF^A_{\k\l} ({\vec\cA}) g^{\m\k} g^{\n\l} \;\; ,\;\;
\cF^2(\cB) \equiv \cF_{\m\n} (\cB) \cF_{\k\l} (\cB) g^{\m\k} g^{\n\l} \; ,
\lab{F2-def} \\
\cF^A_{\m\n} ({\vec\cA}) = 
\pa_\m \cA^A_\n - \pa_\n \cA^A_\m + \eps^{ABC} \cA^B_\m \cA^C_\n \;\; ,\;\;
\cF_{\m\n} (\cB) = \pa_\m \cB_\n - \pa_\n \cB_\m \; .
\lab{F-def}
\er

As shown in Appendix B of Ref.\ct{grav-bags}, for static spherically 
symmetric fields in a static spherically symmetric spacetime metric the 
square-root term $-\h f_0\sqrt{-F^2}$ produces an effective 
{\em ``Cornell''-type confining potential}
\ct{cornell-potential-1,cornell-potential-2,cornell-potential-3}
$V_{\rm eff}(L)$ between charged quantized fermions, $L$ being the distance 
between the latter:
\be
V_{\rm eff} (L) = \sqrt{2}ef_0\; L - \frac{e^2}{2\pi\,L} + 
\bigl( L{\rm -independent} ~{\rm const} \bigr) \; ,
\lab{cornell-type}
\ee
\textsl{i.e.}, $f_0$ and $e$ play the role of a confinement-strength coupling 
constant and of a ``color'' charge, respectively.

In fact, we could equally well take the ``square-root'' nonlinear gauge field 
$\cA_\m$ to be non-Abelian -- for static spherically symmetric solutions 
the non-Abelian model effectively reduces to the abelian one \ct{GG-2}.
Thus, the ``square-root'' gauge field will simulate the QCD-like confining
dynamics.

Now, an important remark is in order. There is a special reason for considering 
precisely the specific form of the non-canonical $f(R)=R+R^2$ gravity-matter 
action \rf{TMMT-1} -- its structure is uniquely fixed by the requirement 
for invariance under global Weyl-scale transformations:
\br
g_{\m\n} \to \l g_{\m\n} \;\; ,\;\; \vp \to \vp + \frac{1}{\a}\ln \l \;\;,\;\; 
\psi \to \l^{-1/2} \psi \;\;,\;\; 
A_{\m\n\k} \to \l A_{\m\n\k} \;\; ,\;\; B_{\m\n\k} \to \l^2 B_{\m\n\k} \;\;,\;\; 
C_{\m\n\k} \to \l C_{\m\n\k} \; ,
\lab{scale-transf} \\
\G^\m_{\n\l} \;,\; H_{\m\n\k} \;,\; \s_a \; ,\; \cA_\m \; ,\;{\vec\cA}_\m \; ,\; \cB_\m \;
- \; {\rm inert} \; .
\nonu
\er

\subsection{3.2 Einstein-Frame Action}
\label{einstein-frame}

The equations of motion of the initial action \rf{TMMT-1} w.r.t. auxiliary 
tensor gauge fields $A_{\m\n\l}$, $B_{\m\n\l}$, $C_{\m\n\l}$ and $H_{\m\n\l}$ 
\be
\pa_\m \Bigl\lb R + L^{(1)}(\vp,\s,\cA_\m,\psi) \Bigr\rb = 0 \;, \;
\pa_\m \Bigl\lb \eps R^2 + L^{(2)}(\vp,\s,\cA_\m,\psi)
+ \frac{\P (H)}{\sqrt{-g}}\Bigr\rb = 0 \;\; ,\;\; \pa_\m \psi^2 = 0 \;\;, \;\; 
\pa_\m \Bigl(\frac{\P_2 (B)}{\sqrt{-g}}\Bigr) = 0 \; ,
\lab{A-B-H-eqs}
\ee
yield the following algebraic constraints:
\be
R + L_1 (\vp,X) + L_2 (\s,Y;\psi) -\h f_0 \sqrt{-F^2} = - M_1 = {\rm const} \; ,
\lab{integr-const-1}
\ee
with $L_1 (\vp,X)$ and $L_2 (\s,Y;\psi)$ as in \rf{L-1a} and \rf{L-1b};
\br
\eps R^2 - \frac{1}{4e^2} F^2 - \frac{1}{4g^2} \cF^2(\vec{\cA}) 
- \frac{1}{4g^{\pr\,2}} \cF^2(\cB) + \frac{\P (H)}{\sqrt{-g}}
- \h g^{\m\n} \pa_\m \psi \pa_\n \psi = - M_2 = {\rm const} \; ,
\lab{integr-const-2}\\
\psi = M_0 = {\rm const} \quad ,\quad 
\frac{\P(B)}{\sqrt{-g}} \equiv \chi_2 = {\rm const} \; ,
\lab{integr-const-3-4}
\er
where $M_0, M_1, M_2$ are arbitrary dimensionful and $\chi_2$ an 
arbitrary dimensionless {\em integration constants}. The algebraic constraint
Eqs.\rf{integr-const-1}-\rf{integr-const-3-4} are the Lagrangian-formalism 
counterparts of the Dirac first-class Hamiltonian constraints on the 
auxiliary tensor gauge fields $A_{\m\n\l},\, B_{\m\n\l},\, H_{\m\n\l}$ 
\ct{grav-bags,grf-essay} (see also \rf{pi-A-const}-\rf{pi-B-pi-H-const} above).

Let us particularly note that the appearance of the dimensionful integration
constants $M_0, M_1, M_2$ signifies a {\em spontaneous breakdown of the global
Weyl scale symmetry} of the starting action \rf{TMMT-1} under \rf{scale-transf}.

The first algebraic constraint in \rf{integr-const-3-4} (the equation of motion
w.r.t. $C_{\m\n\l}$: $\psi = M_0 = {\rm const}$) explains the need to introduce 
the last term in \rf{TMMT-1} with the third non-Riemannian volume element $\P_3 (C)$.
In this way we both preserve the explicit global Weyl-scale invariance of 
\rf{TMMT-1} and dynamically ``freeze'' the second auxiliary scalar field
$\psi$, so that the Higgs-like field $\s$ acquires a dynamically 
generated {\em ordinary} mass term in \rf{L-1b} $L_2 (\s,Y;\psi=M_0) = 
- g^{\m\n}(\nabla_\m \s)^{*}_a \nabla_\n \s_a - M_0^2 \s^{*}_a \s_a$ (before
transferring to the Einstein-frame).

The equations of motion of \rf{TMMT-1} w.r.t. affine connection $\G^\m_{\n\l}$ 
(recall -- we are using Palatini formalism):
\be
\int d^4\,x\,\sqrt{-g} g^{\m\n} \Bigl(\frac{\P_1}{\sqrt{-g}} +
2\eps\,\frac{\P_2}{\sqrt{-g}}\, R\Bigr) \(\nabla_\k \d\G^\k_{\m\n}
- \nabla_\m \d\G^\k_{\k\n}\) = 0 
\lab{var-G}
\ee
yield a solution for $\G^\m_{\n\l}$ as a Levi-Civita connection:
\be
\G^\m_{\n\l} = \G^\m_{\n\l}({\bar g}) = 
\h {\bar g}^{\m\k}\(\pa_\n {\bar g}_{\l\k} + \pa_\l {\bar g}_{\n\k} 
- \pa_\k {\bar g}_{\n\l}\) \; ,
\lab{G-eq}
\ee
w.r.t. to the following {\em Weyl-rescaled metric} ${\bar g}_{\m\n}$:
\be
{\bar g}_{\m\n} = \bigl(\chi_1 + 2\eps\chi_2 R\bigr) g_{\m\n} 
\quad , \quad
\chi_1 \equiv \frac{\P_1 (A)}{\sqrt{-g}} \; ,
\lab{bar-g}
\ee
$\chi_2$ as in \rf{integr-const-3-4}.
Upon using relation \rf{integr-const-1} and notation \rf{integr-const-3-4}
Eq.\rf{bar-g} can be written as:
\be
{\bar g}_{\m\n} = \Bigl\lb\chi_1 - 2\eps\chi_2 \Bigl(L_1(\vp,X) +
L_2(\s,Y;\psi) - \h f_0\sqrt{-F^2} + M_1\Bigr)\Bigr\rb g_{\m\n}\; .
\lab{bar-g-1}
\ee

The Weyl-rescaled metric \rf{bar-g} (or \rf{bar-g-1} is the {\em Einstein-frame
metric} since the corresponding gravity equations of motion of the initial
action \rf{TMMT-1} written in terms of 
${\bar g}_{\m\n}$ acquire the standard form of Einstein equation derivable from an
effective {\em Einstein-frame} action with the canonical Hilbert-Einstein gravity
part w.r.t. ${\bar g}_{\m\n}$ and with the canonical Riemannian
volume element $\sqrt{\det\vert\vert -{\bar g}_{\m\n}\vert\vert}$.

Indeed, as shown in Refs.\ct{bekenstein} the pertinent Einstein-frame
action, where all quantities defined w.r.t. Einstein-frame metric \rf{bar-g}
are indicated by an upper bar, acquires the explicit form:
\be
S = \int d^4 x \sqrt{-{\bar g}} \Bigl\lb R({\bar g}) +
L_{\rm eff} \bigl(\vp,{\bar X};\s,{\bar Y}; {\bar F}^2, 
{\bar\cF}^2(\vec{\cA}),{\bar\cF}^2(\cB)\bigr)\Bigr\rb \; .
\lab{TMMT-einstein-frame}
\ee
Here:
\be
{\bar X} \equiv - \h {\bar g}^{\m\n} \pa_\m \vp \pa_\n \vp \;\; ,\;\;
{\bar Y} \equiv - {\bar g}^{\m\n}(\nabla_\m \s)^{*}_a \nabla_\n \s_a \;\;,
\;\; {\bar F}^2 (\cA) \equiv F_{\m\n} F_{\k\l} {\bar g}^{\m\k} {\bar g}^{\n\l} \; ,
\lab{X-Y-F-bar}
\ee
(and similarly for ${\bar\cF(\vec{\cA})}^2,\, {\bar\cF(\cB)}^2$), and where
the Einstein-frame Lagrangian reads::
\br
L_{\rm eff} = \bigl({\bar X}+{\bar Y}\bigr)\bigl(1-4\eps\chi_2 \cU(\vp,\s)\bigr) +
\eps\chi_2 \bigl({\bar X}+{\bar Y}\bigr)^2 \bigl(1-4\eps\chi_2 \cU(\vp,\s)\bigr)
\nonu \\
- \bigl({\bar X}+{\bar Y}\bigr) \sqrt{-{\bar F}^2} \eps\chi_2\, f_{\rm eff}(\vp,\s) 
- \h f_{\rm eff}(\vp,\s) \sqrt{-{\bar F}^2}
\nonu \\
- \cU(\vp,\s) - \frac{1}{4 e^2_{\rm eff}(\vp,\s)}{\bar F}^2 
-\frac{\chi_2}{4g^2}{\bar\cF}^2(\cA) -\frac{\chi_2}{4g^{\pr\,2}}{\bar\cF}^2(\cB)
\; 
\lab{L-eff-total}
\er
In \rf{L-eff-total} the following notations are used:
\begin{itemize}
\item
$\cU(\vp,\s)$ is the effective scalar field (``inflaton'' + Higgs-like) potential:
\be
\cU(\vp,\s) = \frac{\bigl( f_1 e^{-\a\vp} + M^2_0 \s^{*}\s - M_1\bigr)^2}{4\chi_2
\bigl\lb M_2 + \eps \bigl( f_1 e^{-\a\vp} + M^2_0 \s^{*}\s - M_1\bigr)^2\bigr\rb} \; .
\lab{U-vp-s}
\ee
\item
$f_{\rm eff}(\vp,\s)$ is the effective confinement-strength coupling constant:
\be
f_{\rm eff}(\vp,\s) = f_0 \bigl(1-4\eps\chi_2 \cU(\vp,\s)\bigr) \; ;
\lab{f-eff}
\ee
\item
$e^2_{\rm eff}(\vp,\s)$ is the effective ``color'' charge squared:
\be
e^2_{\rm eff}(\vp,\s) = \frac{e^2}{\chi_2}
\Bigl\lb 1 + \eps e^2 f_0^2 \bigl(1-4\eps\chi_2 \cU(\vp,\s)\bigr) \Bigr\rb^{-1}
\lab{e-eff}
\ee
\end{itemize}

Note that \rf{L-eff-total} is of quadratic {\em ``k-essence''} type
\ct{k-essence-1,k-essence-2,k-essence-3,k-essence-4} w.r.t.
``inflaton'' $\vp$ and the Higgs-like $\s$ fields.

\section{4. Cosmological Implications}
\label{cosmolog}


The nonlinear ``confining'' gauge field $\cA_\m$ develops a nontrivial vacuum
field-strength:
\be
\frac{\pa L_{\rm eff}}{\pa {\bar F}^2}\bgv_{{\bar X},{\bar Y}=0} = 0
\lab{F-vac-eq}
\ee
explicitly given by:
\be
\sqrt{-{\bar F}^2}_{\rm vac} = f_{\rm eff}(\vp,\s)\, e^2_{\rm eff}(\vp,\s)
\lab{F-vac}
\ee
Substituting \rf{F-vac} into \rf{L-eff-total} we obtain the following total
effective scalar potential (with $\cU(\vp,\s)$ as in \rf{U-vp-s}):
\be
\cU_{\rm total}(\vp,\s) = \frac{\cU(\vp,\s)(1-\eps e^2 f_0^2) + 
e^2 f_0^2/4\chi_2}{1 + \eps e^2 f_0^2 \bigl(1-4\eps\chi_2 \cU(\vp,\s)\bigr)} \; .
\lab{U-eff-total}
\ee
$\cU_{\rm total}(\vp,\s)$ \rf{U-eff-total} has few remarkable properties.
First, $\cU_{\rm total}(\vp,\s)$ possesses two infinitely large flat regions
as function of $\vp$ when $\s$ is fixed:

(a) (-) flat ``inflaton'' region for large negative values of $\vp$;

(b) (+) flat ``inflaton'' region for large positive values of $\vp$ with 
$\s$ fixed;
\\
respectively, as depicted on Fig.1 (for $M_0 \s^{*}\s \leq M_1$) or 
Fig.2 (for $M_0 \s^{*}\s \geq M_1$).

\begin{figure}
\includegraphics[width=9cm,keepaspectratio=true]{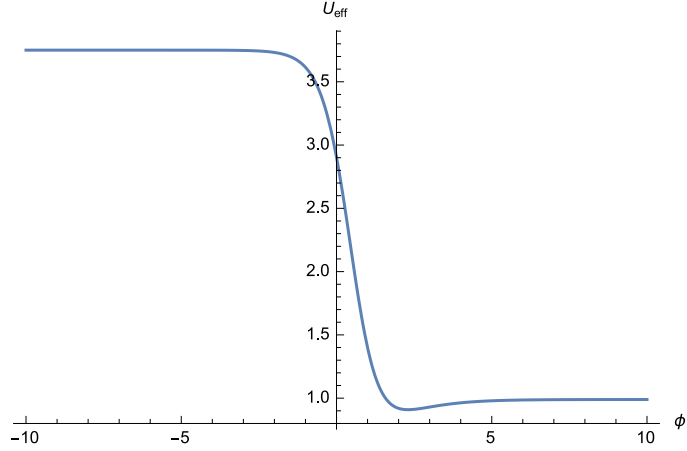}
\caption{Qualitative shape of the total effective scalar potential $U_{\rm total}$ 
\rf{U-eff-total} as function of the ``inflaton''$\vp$ for fixed Higgs-like $\s$
(when $M_0 \s^{*}\s \leq M_1$).}
\end{figure}

\begin{figure}
\includegraphics[width=9cm,keepaspectratio=true]{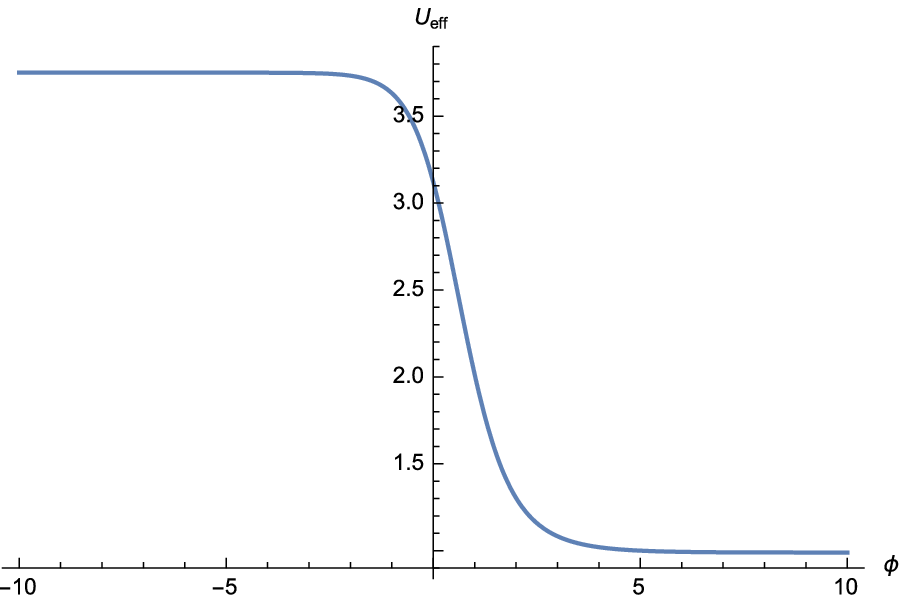}
\caption{Qualitative shape of the total effective scalar potential $U_{\rm total}$ 
\rf{U-eff-total} as function of the ``inflaton''$\vp$ for fixed Higgs-like $\s$
(when $M_0 \s^{*}\s \geq M_1$).}
\end{figure}

In the (-) flat ``inflaton'' region the effective scalar field potential 
reduces to:
\be
\cU(\vp,\s ={\rm fixed}) \simeq \frac{1}{4\eps\chi_2}\quad\longrightarrow\quad
\cU_{\rm total} \simeq \cU^{(-)}_{\rm total} = \frac{1}{4\eps\chi_2} \; ,
\lab{U-minus}
\ee
implying that all terms containing $\vp$ and $\s$ disappear from the
Einstein-frame Lagrangian \rf{TMMT-einstein-frame}. Thus, there is no
$\s$-field potential and, therefore,
{\em no electroweak spontaneous breakdown in the (-) flat ``inflaton'' region}. 

From \rf{f-eff} the first relation \rf{U-minus} implies $f_{\rm eff} = 0$. 
Recalling that $f_{\rm eff}$, the effective coupling constant of the
square-root Maxwell term, measures the charge-confining strength according to
\ct{GG-2,grav-bags}, we conclude that there is 
{\em confinement is suppressed in the (-) flat ``inflaton'' region}.

In the (+) flat ``inflaton'' region
the effective scalar field potential becomes:
\br
\cU(\vp,\s) \simeq \cU_{(+)}(\s) = \frac{\bigl(M_0^2 \s^{*}\s - M_1\bigr)^2}{
4\chi_2 \bigl\lb M_2 + \eps \bigl(M_0^2 \s^{*}\s - M_1\bigr)^2\bigr\rb}
\lab{U-plus} \\
\longrightarrow \quad 
\cU_{\rm total} (\vp,\s) \simeq \cU^{(+)}_{\rm total}(\s) = 
\frac{\cU_{(+)}(\s)(1-\eps e^2 f_0^2) + e^2 f_0^2/4\chi_2}{1
+ \eps e^2 f_0^2 \bigl(1-4\eps\chi_2 \cU_{(+)}(\s)\bigr)}
\lab{U-total-plus}
\er
producing a dynamically generated {\em nontrivial vacuum for the Higgs-like field}:
\be
|\s_{\rm vac}|= \sqrt{M_1}/M_0 \; ,
\lab{higgs-vac}
\ee
\textsl{i.e.}, we obtain {\em ``gravity-inflaton-assisted'' electroweak spontaneous breakdown
in the (+) flat ``inflaton'' region}.

At the Higgs vacuum we have dynamically generated vacuum energy density
(cosmological constant):
\be
\cU^{(+)}_{\rm total}(\s_{\rm vac}) \equiv 2 \L_{(+)} = \eps e^2 f_0^2
\Bigl\lb 4\eps\chi_2 \bigl( 1 + \eps e^2 f_0^2\bigr)\Bigr\rb^{-1} \; .
\lab{CC-plus}
\ee

The effective confinement-strength coupling constant:
\be
f_{\rm eff} \simeq f_{(+)} = f_0 \bigl(1-4\eps\chi_2 \cU_{(+)}(\s)\bigr) > 0\; ,
\lab{f-plus}
\ee
therefore we obtain {\em ``gravity-inflaton-assisted'' charge confinement in the 
(+) flat ``inflaton'' region}.

As seen from Fig.1 or Fig.2, the heights of the two flat ``inflaton'' regions 
of the total scalar potential, \textsl{i.e.}, the corresponding vacuum
energies are given by \rf{U-minus} and \rf{CC-plus}, respectively:
\be
\cU^{(-)}_{\rm total} = \frac{1}{4\eps\chi_2} \quad ,\quad
\cU^{(+)}_{\rm total}(\s_{\rm vac}) \equiv 2 \L_{(+)} = \eps e^2 f_0^2
\Bigl\lb 4\eps\chi_2 \bigl( 1 + \eps e^2 f_0^2\bigr)\Bigr\rb^{-1} \; .
\lab{}
\ee
Thus, the $(-)$ and the $(+)$ flat ``inflaton'' regions of the effective
``inflaton'' potential with a very large
and a very small height, respectively, can be accordingly
identified as describing the ``early'' (``inflationary'') and ``late'' 
(today's dark energy dominated) epoch of the universe provided we take the 
following numerical values for the parameters in order to conform to the 
{\em PLANCK} data \ct{Planck-1,Planck-2}:
\be
\cU^{(-)}_{\rm total} \sim 10^{-8} M_{\rm Pl}^4 \to 
\eps\chi_2 \sim 10^8 M_{\rm Pl}^{-4}
\;\; ,\;\; \L_{(+)} \sim 10^{-122} M_{\rm Pl}^4 \to 
\frac{e^2 f_0^2}{\chi_2} \sim 10^{-122} M_{\rm Pl}^4 \; ,
\lab{param-1}
\ee
where $M_{\rm Pl}$ is the Planck mass scale.

From the Higgs v.e.v. $|\s_{\rm vac}|= \sqrt{M_1}/M_0$ and the Higgs mass
$\frac{M_1 M_0^2}{4\chi_2 M_2}$ resulting from the dynamically generated
Higgs-like potential $\cU^{(+)}_{\rm total}(\s)$ \rf{U-total-plus} we find:
\be
M_0 \sim M_{\rm EW} \;\; ,\;\; M_{1,2}\sim M_{\rm EW}^4 \; ,
\lab{param-2}
\ee
where $M_{\rm EW} \sim 10^{-16} M_{\rm Pl}$ is the electroweak mass scale.

\section{5. Conclusions and Outlook}
\label{conclude}

Here we have proposed a non-canonical model of $f(R)=R+R^2$ gravity coupled
to the ``inflaton'' and
the bosonic part of the standard particle model, incorporating two main building 
blocks -- employing the formalism of non-Riemannian spacetime volume forms
(generally covariant metric-independent volume elements) as well as introducing 
a special strongly non-linear gauge field with a square-root of the usual 
Maxwell/Yang-Mills kinetic term simulating QCD-like confinement dynamics. 
Due to the special interplay of the dynamics of the above principal
ingredients our model is capable of producing in the Einstein frame:

\begin{itemize}
\item
Unified ``quintessential'' description of the evolution of the ``early''
and ``late'' universe due to a natural dynamical generation of vastly
different vacuum energy densities thanks to the auxiliary non-Riemannian
volume-form antisymmetric tensor gauge fields;
\item
gravity-inflaton-assisted dynamical generation of Higgs-like electroweak spontaneous
symmetry breaking effective scalar potential in the ``late'' universe, as
well as gravity-inflaton-assisted charge confinement mechanism through the
``square-root'' nonlinear gauge field;
\item
Gravity-inflaton-induced suppression of electroweak spontaneous symmetry breaking, as
well as gravity-inflaton-induced deconfinement in the ``early'' universe.
\item
Apart from the cosmological implications discussed above, 
the non-Riemannian volume-form formalism has further physically relevant
applications such as producing a novel mechanism for supersymmetric
Brout-Englert-Higgs effect in supergravity through dynamical generation of a
cosmological constant, which triggers spontaneous supersymmetry breaking and
dynamical gravitino mass generation \ct{susyssb-1,susyssb-2}.
\end{itemize}

Let us also note that the QCD-simulating ``square-root'' nonlinear gauge field when
interacting with gravity produces several other interesting effects:

(a) black holes with an additional constant
background electric field exercising confining force on charged test
particles even when the black hole itself is electrically neutral
\ct{grav-cornell}; 

(b) Coupling to a charged lightlike brane produces a charge-``hiding'' lightlike
thin-shell wormhole, where a genuinely charged matter source is detected as 
electrically neutral by an external observer \ct{hide-confine}.

(c) Coupling to two oppositely charged lightlike brane sources produces a 
two-``throat'' lightlike thin-shell wormhole displaying a genuine QCD-like 
charge confinement, \textsl{i.e.}, the whole electric flux is trapped within
a tube-like spacetime region connected the two charged lightlike branes
\ct{hide-confine}.

(d) Charge confining gravitational electrovacuum shock wave \ct{shock}.

The present model needs some further improvements. First of all 
it is necessary to avoid getting 
an unnaturally small value for the effective confinement strength coupling 
constant $f_0$ in the ``late'' universe resulting from the second relation 
\rf{param-1} (the latter was needed for compatibility with the {\em PLANCK} data 
\ct{Planck-1,Planck-2} for the value of today's cosmological constant). 

Further important task must be the inclusion of the
fermions in order to incorporate more faithfully the full standard particle model.
To this end we can follow the steps outlined in several previous papers by
some of us \ct{fermion-families,DE-DM-fermions,neutrino-DE} 
devoted to the study of modified gravity within the
non-Riemannian volume element formalism coupled to fermionic matter fields.
 




\section{ACKNOWLEDGMENTS}
We gratefully acknowledge support of our collaboration through 
the academic exchange agreement between the Ben-Gurion University in Beer-Sheva,
Israel, and the Bulgarian Academy of Sciences. 
E.N. and E.G. have received partial support from European COST actions
MP-1405 and CA-16104, and from CA-15117 and CA-16104, respectively.
E.N. and S.P. are also thankful to Bulgarian National Science Fund for
support via research grant DN-18/1. 



\begin{thebibliography}{9}
\bibitem{general-cit-1}
E. Kolb and M. Turner, {\em ``The Early Universe''} (Addison Wesley, 1990).
\bibitem{general-cit-2}
A. Linde, {\em ``Particle Physics and Inflationary Cosmology''},
(Harwood, Chur, Switzerland, 1990).
\bibitem{general-cit-3}
A. Guth, {\em ``The Inflationary Universe''} (Addison-Wesley, 1997).
\bibitem{general-cit-4}
A. Liddle and D. Lyth, {\em ``Cosmological Inflation and Large-Scale Structure''}
(Cambridge Univ. Press, 2000).
\bibitem{general-cit-5}
S. Dodelson, {\em ``Modern Cosmology''} (Acad. Press, 2003).
\bibitem{general-cit-6}
V. Mukhanov, {\em ``Physical Foundations of Cosmology''} 
(Cambridge Univ. Press, 2005).
\bibitem{general-cit-7}
S. Weinberg, {\em ``Cosmology''} (Oxford Univ. Press, 2008).
\bibitem{susyssb-1}
E. Guendelman, E. Nissimov and S. Pacheva, \textsl{Bulg. J. Phys.} {\bf 41}, 
123 (2014) ~(\textsl{arXiv:1404.4733}).
\bibitem{grav-bags}
E. Guendelman, E. Nissimov and S. Pacheva, \IJMPA{30}{2015}{1550133} 
~(\textsl{arXiv:1504.01031}).
\bibitem{TMT-orig-1}
E. Guendelman, \textsl{Mod. Phys. Lett.} {\bf A14} 1043-1052 (1999)
~(\textsl{arXiv:gr-qc/9901017}).
\bibitem{TMT-orig-2}
E. Guendelman and A. Kaganovich,
\textsl{Phys. Rev.} {\bf D60} 065004 (1999) ~(\textsl{arXiv:gr-qc/9905029}).
\bibitem{grf-essay}
E. Guendelman, E. Nissimov and S. Pacheva, \IJMPD{25}{2016}{1644008}
~(\textsl{1603.06231}).
\bibitem{gravity-assist-86}
J. Bekenstein, \textsl{Found. Phys.} {\bf 16}, 409 (1986).
\bibitem{GG-2}
P. Gaete and E. Guendelman, {\sl Phys. Lett.} {\bf B640} 201-204 (2006)
~(\textsl{arXiv:hep-th/0607113}).
\bibitem{thooft}
G. 't Hooft, \textsl{Nucl. Phys. B (Proc. Suppl.)} {\bf 121} 333-340 (2003)
~(\textsl{arXiv:0208054}[hep-th]).
\bibitem{bekenstein}
E. Guendelman, E. Nissimov and S. Pacheva, in {\em Jacob Bekenstein Memorial
Volume} (World Scientific, 2018), to be published, (\textsl{arXiv:1804.07925}.
\bibitem{cornell-potential-1}
E. Eichten, K. Gottfried, T. Kinoshita, J. Kogut, K. Lane and T.-M. Yan, 
\PRL{34}{1975}{369-372}.
\bibitem{cornell-potential-2}
W. Buchm{\"u}ller (ed.), {\em ``Quarkonia''}, \textsl{Current Physics Sources and 
Comments}, vol.9, North Holland (1992).
\bibitem{cornell-potential-3}
M. Karliner, B. Keren-Zur, H. Lipkin and J. Rosner, \AoP{324}{2009}{2-15}
~(\textsl{0804.1575}[hep-ph]).
\bibitem{k-essence-1}
T. Chiba, T.Okabe and M. Yamaguchi, \PRD{62}{2000}{023511} 
~(\textsl{arXiv:astro-ph/9912463}).
\bibitem{k-essence-2}
C. Armendariz-Picon, V. Mukhanov and P. Steinhardt, \PRL{85}{2000}{4438}
~(\textsl{arXiv:astro-ph/0004134}).
\bibitem{k-essence-3}
C. Armendariz-Picon, V. Mukhanov and P. Steinhardt, 
\PRD{63}{2001}{103510} ~(\textsl{arXiv:astro-ph/0006373}).
\bibitem{k-essence-4}
T. Chiba, \PRD{66}{2002}{063514} ~(\textsl{arXiv:astro-ph/0206298}).
\bibitem{Planck-1}
R. Adam {\it et al.} (Planck Collaboration), \textsl{Astron. Astrophys.} 
{\bf 571}, A22 (2014) ~(\textsl{arXiv:1303.5082} [astro-ph.CO]).
\bibitem{Planck-2}
R. Adam {\em et al.} (Planck Collaboration), \textsl{Astron. Astrophys.}
{\bf 586}, A133 (2016) ~(\textsl{arXiv:1409.5738} [astro-ph.CO]).
\bibitem{susyssb-2}
E. Guendelman, E. Nissimov and S. Pacheva, in \textsl{Eight Mathematical
Physics Meeting}, ed. by B. Dragovich and I. Salom (Belgrade Inst. Phys.
Press, 2015), pp. 105-115 ~(\textsl{arXiv:1501.05518}).
\bibitem{grav-cornell}
E. Guendelman, E. Nissimov and S. Pacheva, \PLB{704}{2011}{230}, 
erratum \PLB{705}{2011}{545} ~(\textsl{arXiv:1108.0160}).
\bibitem{hide-confine}
E. Guendelman, E. Nissimov and S. Pacheva, \IJMPA{26}{2011}{5211}
~(\textsl{arXiv:1109.0453}).
\bibitem{shock}
E. Guendelman, E. Nissimov and S. Pacheva, \MPLA{29}{2014}{1450020}
~(\textsl{arXiv:1310.1558}).
\bibitem{fermion-families}
E. Guendelman and A. Kaganovich, \MPLA{17}{2002}{1227} 
~(\textsl{arXiv:hep-th/0110221}).
\bibitem{DE-DM-fermions}
E. Guendelman and A. Kaganovich, \IJMPA{19}{2004}{5325}
~(\textsl{arXiv:gr-qc/0408026}).
\bibitem{neutrino-DE}
E. Guendelman and A. Kaganovich, \IJMPA{21}{2006}{4373}
~(\textsl{arXiv:gr-qc/0603070}).
\end{thebibliography}
\end{document}